\documentstyle[epsfig]{l-aa}
\def\arcsper  {\ifmmode \rlap.{'' }\else $\rlap{.}'' $\fi}
\def\arcmper  {\ifmmode \rlap.{' }\else $\rlap{.}' $\fi}
\def\rdv{${\rm r^{1/4}}$}
\def\kms{\ifmmode{{\rm ~km~s}^{-1}}\else{~km~s$^{-1}$}\fi}
\def\simlt       {\lower.5ex\hbox{$\; \buildrel < \over \sim \;$}}
\def\amm      {\AA\thinspace mm$^{-1}$}    
\def\arcs   {$^{\prime\prime}$}
\def\>           {$>$}
\def\<           {$<$}
\def\approx   {$\sim$}
\def\deg      {\ifmmode^\circ\else$^\circ$\fi}   
\def\solmass  {M$_\odot$}
\def\solum    {L$_\odot$}
\begin{document}
\thesaurus{03 (11.03.4 Hickson 96; 11.09.2; 11.11.1; 11.05.2; 11.19.6)}
\title{Hickson 96: a physical compact group \thanks{Based partially on data
obtained at the Observatoire de Haute Provence, France}}
\author{L. Verdes--Montenegro \inst{1}
\and A. del Olmo \thanks{Visiting Astronomer German-Spanish Astronomical 
Center, Calar Alto, operated by the Max-Planck-Institute for Astronomy, 
Heidelberg, jointly with the Spanish National Commission for Astronomy}
\inst{1} \and J. Perea$^{\star\star}$\inst{1} \and 
E. Athanassoula\inst{2} \and I. M\'{a}rquez\inst{1} \and R. Augarde\inst{2}}
\institute{Instituto de Astrof\'{\i}sica de Andaluc\'{\i}a, CSIC,
Apdo. 3004, 18080 Granada, Spain \and Observatoire de Marseille, 2 Place
le Verrier, 13248 Marseille Cedex 4, France.}
\offprints{L. Verdes-Montenegro}
\date{Received {23 August 1996}/ Accepted {22 October 1996}} \maketitle

\begin{abstract}

We analyze the morphology and dynamics of the galaxies of the Hickson 96
compact group by means of deep CCD images in the B, V and R photometric
bands and long--slit spectroscopy. 

The four galaxies of this spiral rich group show signs of gravitational
interaction. Two long tails come out from the contact region of the close
pair H96ac. It is also there where both galaxies, bi-symmetrical in their
inner parts, loose one of their arms. Moreover, both the photometric and
kinematical center of H96a are displaced relative to the center of the
disk. 

H96b, a giant elliptical galaxy, shows significant deviations from a \rdv
law at the inner parts and small rotation along its major axis. We find at
the centre of the galaxy an elongated component which is kinematically
decoupled.  A wide faint plume seems to emerge from this galaxy.  H96d, the
smallest galaxy of the group, seems to be influenced by the bigger members,
as indicated by three prominent knots of recent bursts of star formation in
its blue disk, and by hints of optical bridges joining H96d with H96a and
b. 

Finally the group has a low velocity dispersion (160\kms) and is well
isolated - no galaxies with comparable magnitude to H96a or H96b are found
in its neighborhood.  All these results lead us to conclude that Hickson 96
constitutes a real physical system. 


\end{abstract}

\section{Introduction}

Hickson groups are among the densest galaxy systems in the Universe, with
apparent surface density enhancements that range from 300 to 2000 (Sulentic
1987).  They are defined by the number of members, (n $\geq$ 4),
compactness and isolation degree (Hickson 1982, 1993).  The original
subsample has been enlarged in the southern sky by means of an automated
search by Prandoni et al. (1994) and relaxing Hickson criteria (Tassi \&
Iovino, 1995).  The small number of members in Hickson groups allows a
detailed study of individual galaxies. In addition, and contrary to other
compact groups such as those of Shakbazyan (Del Olmo, Moles \& Perea,
1995), Hickson groups exhibit a wide range of morphologies and degrees of
interaction. This variety is quite different from that of other high
density regions such as the centers of rich clusters. 

Their high surface density enhancements, together with low velocity
dispersions ($<$ $\sigma$ $>$$_{2D}$ = 200 \kms , $<$ $\sigma$ $>$$_{3D}$ =
330 \kms \, Hickson et al. 1992), argue that the groups are physical but
with short dynamical lifetimes (\simlt$10^9$ yrs). The number of merger
candidates, however, appears to be extremely low (Zepf et al. 1991; Moles
et al. 1994) and, although star formation is enhanced with respect to
isolated galaxies, it is lower than in strongly interacting pairs (Moles et
al.  1994; Sulentic \& Rabaca 1994).  A diffuse medium surrounding entire
groups has been detected as atomic gas (Williams \& Van Gorkom 1995, and
references therein), X-ray emission (Bahcall et al. 1984; Ponman \& Bertran
1993; Ebeling et al.  1994; Saracco \& Ciliegi 1995; Mulchaey et al. 1996;
Pildis et al.  1995a; Sulentic et al. 1995), and diffuse optical light
(Pildis 1995; Pildis et al. 1995b; Sulentic 1987, and references therein).
The study of Hickson group environments shows that some of them are
embedded in more extended physical systems (Ramella et al. 1994; Rood \&
Struble 1994), but with low densities and in general well isolated
(Sulentic 1987; Rood \& Williams 1989). 
\begin{figure*}
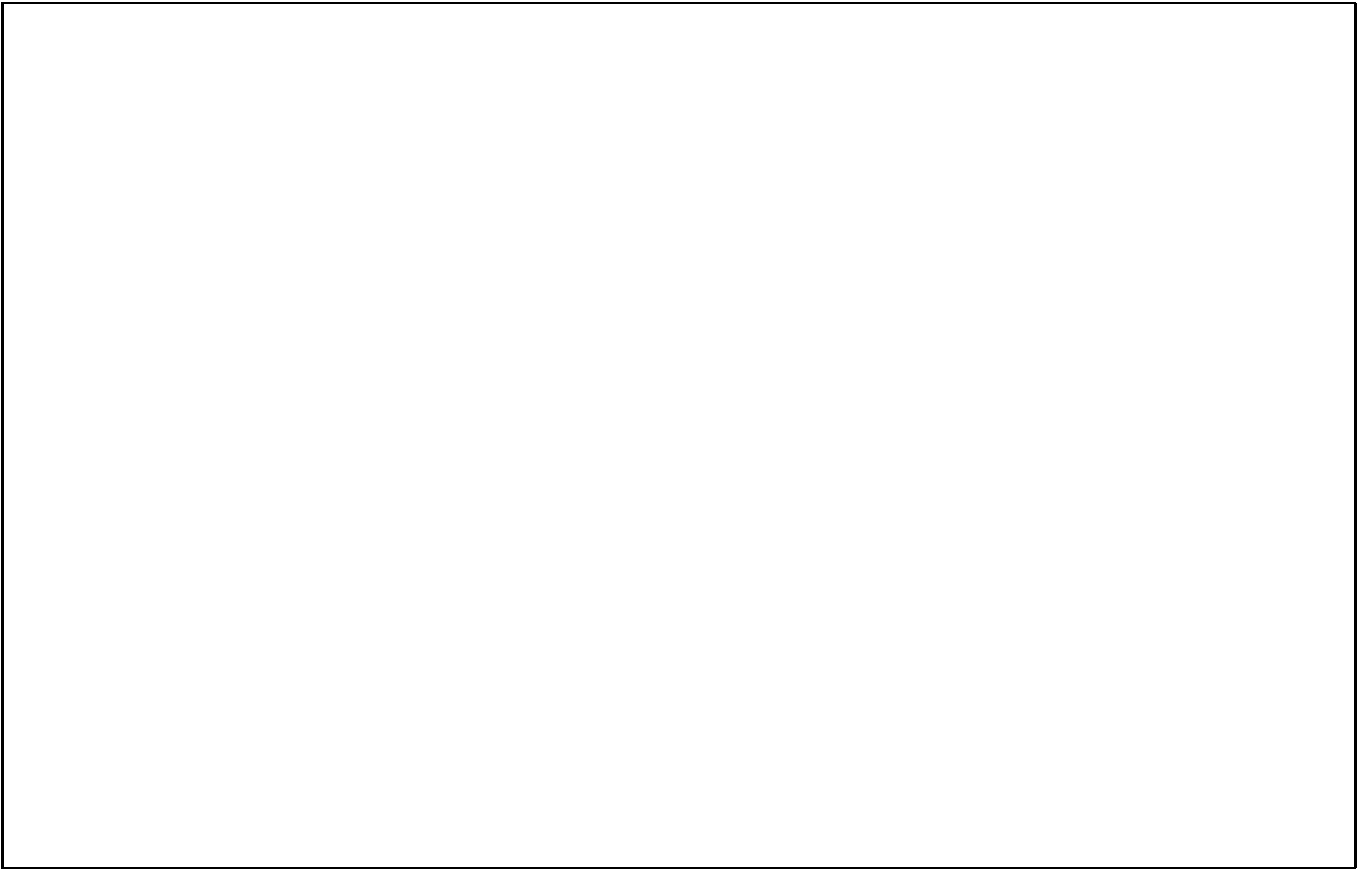

\picplace{11.5cm}
\caption[ ]{V-band image of Hickson 96 in a logarithmic grey 
scale representation. Higher intensities are
darker. The orientation of all the images is North up and East to 
the left. Positions are expressed as offsets with respect to the center of
H96a.}
\end{figure*}

Different models have been proposed in order to explain the nature of
Hickson groups. Mamon (1986, 1995) suggested that 50\% of the groups are
chance superposition of pairs and unrelated galaxies within loose groups. 
Hernquist et al. (1995) interpret them as end-on views of filaments of
galaxies.  In the context of the last model, and based on X-ray results by
Mulchaey et al. (1996), Ostriker et al. (1995) suggest that low spiral
fraction groups are real physical entities, but high spiral fraction groups
are projected fi\-la\-ments.  The application of their $Q$ parameter to
Hickson groups has been found however inappropriate by Pildis et al. (1996)
because the groups are gas poor. Dell'Antonio et al. (1995) have also
argued that the large axial ratios inferred by Ostriker et al. (1995) could
be just a reflection of the difference in gas fraction between groups and
the reference rich clusters.  Diaferio et al. (1994) proposed that compact
groups could continually form in rich groups, while Athanassoula et al.
(1996) find that, if they have a sufficiently massive and not too centrally
concentrated common dark halo, and/or appropriate initial kinematics, their
lifetimes should be considerably longer than would be naively expected. 
Since no significant enhancement in the SFR is observed nor many candidates
for mergers of late types galaxies, the effects of the environment on the
gaseous and stellar component of Hickson groups could appear only at faint
levels in which case detailed observations should be needed to reveal them.
 In order to bring new elements to this debate we will discuss here one
Hickson compact group with a high spiral fraction, Hickson 96. 

Our aim is to look for optical signatures of interactions by studying the
dynamics of the interaction - both in terms of the overall appearance of
the group and the local effects on member galaxies - and the possible
perturbations of the star formation activity.  We report here deep
photometric data and spectroscopic observations for the four galaxies of
the group.  In $\S$2 we describe the observations and data reduction, in
$\S$3 we show our results for the group as a whole and for the individual
galaxies, and we discuss them and present our main conclusions in $\S$4. 

A Hubble constant $\rm{H_0=100\,km\,s^{-1}Mpc^{-1}}$ is used throughout
this paper. 

\section{Observations and data reduction}
\subsection{Surface Photometry}

We obtained CCD images of Hickson 96 in the Johnson BVR bands.  Fig. 1
shows the V band image of the entire group. 
\begin{table*}
\caption[ ]{Long slit spectra.}

\begin{tabular}{cccccc}
\hline\noalign{\smallskip}
Galaxies& Date of Observation & Dispersion &  Exposure Time  & 
Spectral Range & Position Angle
\\
 &  & \amm & (s) & \AA & ($^{\circ})$\\
\noalign{\smallskip}
\hline\noalign{\smallskip}
a - c & 5-11-91 & 33 & 5400 & 6600 - 7100 & 66 \\
b - d & 6-11-91 & 260& 3600 & 3650 - 7350 & 88 \\
a - c & 6-11-91 & 260& 3600 & 3650 - 7350 & 66 \\
c - d & 8-11-91 & 260& 5400 & 3650 - 7350 & 161 \\
b major& 16-6-93 & 72 & 2820 & 4425 - 6335 & 53\\
b minor& 17-6-93 & 72 & 2220 & 4425 - 6335 & 143\\
\noalign{\smallskip}
\hline
\end{tabular}
\end{table*}
Offset positions are in arcsec relatives to the center of NGC 7674 $\alpha
(1950) = 23^{h}25^{m}24.^{s}4$, $\delta (1950) = 08^{\circ} 30^{\prime}
13^{\prime\prime}$.  The images were obtained at the prime focus of the
3.5m telescope of the Centro Astron\'{o}mico Hispano-Alem\'{a}n in Calar
Alto (Spain). The detector was an RCA CCD of 1024$\times$640 pixels of
$15\mu m$ size, giving a scale of 0\arcsper 25 pixel$^{-1}$, and a field of
view of 4\arcmper 2 $\times$ 2\arcmper 4. The exposure times were 1000s,
700s and 500s in B, V and R respectively.  The seeing of each frame was
measured from the FWHM obtained by fitting an analytic Moffat function to
the profiles of stars in the field of the group. It was found to be
1\arcsper 0 in B, 0\arcsper 8 in V and 1\arcsper 0 in R.

The reduction and calibration of the data were carried out using standard
techniques and produced fluxes that are accurate at the 2\% level.  Bias
exposures taken through the run were found to be constant and were used to
construct an average bias which was subtracted from each image.
Pixel-to-pixel variations were evaluated with dome flat-fields.  All flats,
after normalization by their median values, show similar form and no
important diffe\-rences.  Images were divided by the averaged flat in each
filter. The atmospheric extinction was determined from observations of
three selected fields in the open clusters NGC 272, Be87 and NGC 366. The
stars were chosen to cover a wide range of colours to take into account
colour effects in the standard system. The rms errors of the standard stars
in the final calibration are smaller than 0.05 mag in all colours. We
subtracted a constant sky background from each frame.  The errors due to
variations in the sky were always smaller than 1\%. Colour indexes given in
this paper and noted with subindex $0$, have been corrected for galactic
absorption (using the extinction value given by Burstein \& Heiles, 1984,
with the reddening law from Savage \& Mathis, 1979), internal extinction
(de Vaucouleurs et al., RC3, 1991) and K-effect. 

We have compared our photometric data with those by Longo \& de Vaucouleurs
(1983) and Hickson, Kindl \& Auman (1989). V magnitudes are consistent to
better than 2\% and colour indexes to better than 8\% for NGC 7675.
Agreements are better than 0.1 mag for apertures smaller than 55\arcs\, for
NGC 7674. At larger radii the photoelectric photometry is probably affected
by the close companion H96c and a star. 

\subsection {Spectroscopy}

Table 1 summarizes the long slit spectra taken for this study.  The format
is as follows: Column 1) spectrum identification; 2) date of observation;
3) spectral dispersion; 4) exposure time in seconds; 5) spectral range and
6) position angle of the slit in degrees. The first four spectra were taken
with the reflective aspherized grating spectrograph Carelec (Lema\^{\i}tre
et al. 1990) used at the Cassegrain focus of the 193 cm telescope at the
Observatoire de Haute Provence.  The first of them was obtained with a
thick front illuminated Thomson CCD with 576$\times$384 pixels of 23$\mu m$
size. The reciprocal dispersion of 33\AA \, mm$^{-1}$ gives 0.76 \AA \,
pixel$^{-1}$ and a spectral resolution of 1.9 \AA.  The spatial scale is
1\arcs \, pixel$^{-1}$. The slit passed through the centers of H96a - c. 
The next three spectra (through the centers of H96b - d, a - c, and c - d)
were taken with a thinned back illuminated RCA CCD with 512$\times$320
pixels. The size of a pixel was 30$\mu m$ and the dispersion was 260\AA \,
mm$^{-1}$ yielding 7.8\AA \, pixel$^{-1}$ and a spectral resolution of 15.6
\AA. The spatial scale is 1\arcsper 3 pixel$^{-1}$.  The slit width of
2\arcsper 5 provided good sampling with the seeing at Haute Provence during
that run.  The spectra were reduced using the usual methods including the
instrumental correction by spectrophotometric standard stars observed each
night. Suitable procedures were written inside the ESO MIDAS package. 

Two additional long-slit spectra were taken along the minor and major axis
of NGC 7675 at the 3.5 m telescope on Calar Alto, using the Twin
spectrograph at the Cassegrain focus. As detector we use a TEK CCD camera
with 1024 $\times$ 1024 pixels of 24$\mu m$.  The spatial resolution was
0\arcsper 9 pixel$^{-1}$. We use a slit width of 250$\mu m$, corresponding
to 1.5\arcs . The observing conditions were good, with a seeing better than
1\arcsper 2. In this run we also obtained a total of 15 exposures of
standard radial velocity giant stars of G and K type, in order to obtain
accurate redshifts and velocity dispersions.  These were calculated using
the cross--correlation technique derived by Tonry \& Davis (1979).  The
wavelength calibration was tested using sky lines present in the spectra.
The rms of the central wavelengths in the lines for both spectra was less
than 0.008\AA, corresponding to 2\kms. We show one low resolution spectrum
for each galaxy in Fig. 2. 
\begin{figure*}
\epsfig{file=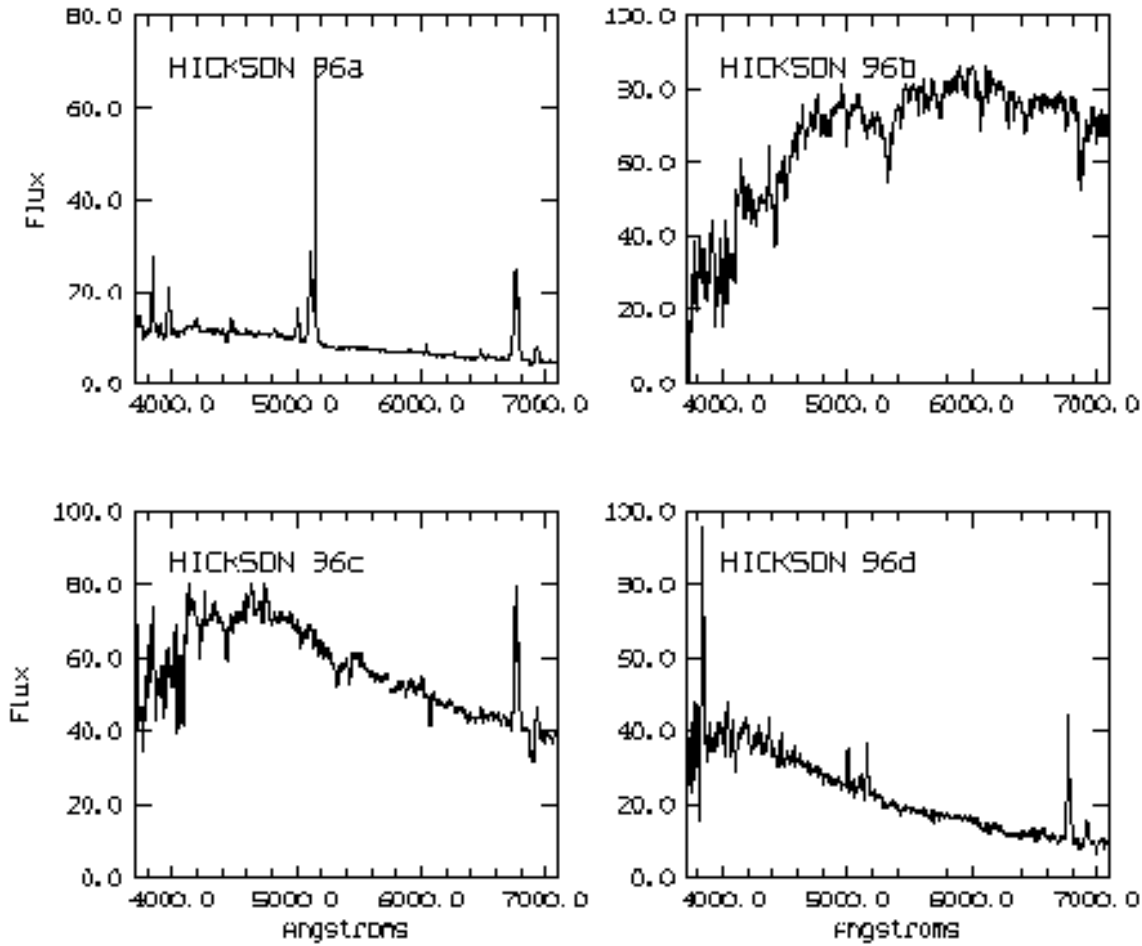,height=14truecm}
\caption[ ]{Spectra of galaxies H96a, H96b, H96c  and H96d 
with position angles of respectively 66$^{\circ}$, 88$^{\circ}$, 
66$^{\circ}$ and 161$^{\circ}$.}
\end{figure*}
\begin{table}
\caption[ ]{Velocities of HCG 96 members.}

\begin{tabular}{cc}
\hline\noalign{\smallskip}
Galaxy& Heliocentric velocity\\
&  (km s$^{-1}$) \\
\noalign{\smallskip}
\hline
\noalign{\smallskip}
H96a&8670 $\pm$ 20\\
H96b&8570 $\pm$ 50\\
H96c&8800 $\pm$ 20\\
H96d&9000 $\pm$ 70\\
\noalign{\smallskip}
\hline
\end{tabular}
   \end{table}

In order to calculate the heliocentric velocity for each galaxy we have
used the high resolution spectra for H96a and c, and compared the results
with the values obtained for the low resolution spectra. In the case of
H96b we have used a mean of the values obtained from the 3 different
available slits (major axis, minor axis and p. a. = 88$^{\circ}$). For H96d
we only have a low resolution spectrum. For the emission lines we assume
that the center of each measured spectral line coincides with the center of
the best Gaussian fit.  For several cases, where H$\alpha$, and [NII] or
[OIII] were blended due to the low resolution, we have separated the
components by a multiple gaussian fit made with a routine included in
MIDAS.  The results of these measures are given in Table 2. There is a good
agreement between our values and those obtained by Hickson et al. (1992). 

\section{Results}
\subsection{The group and its environment}
\begin{figure*}
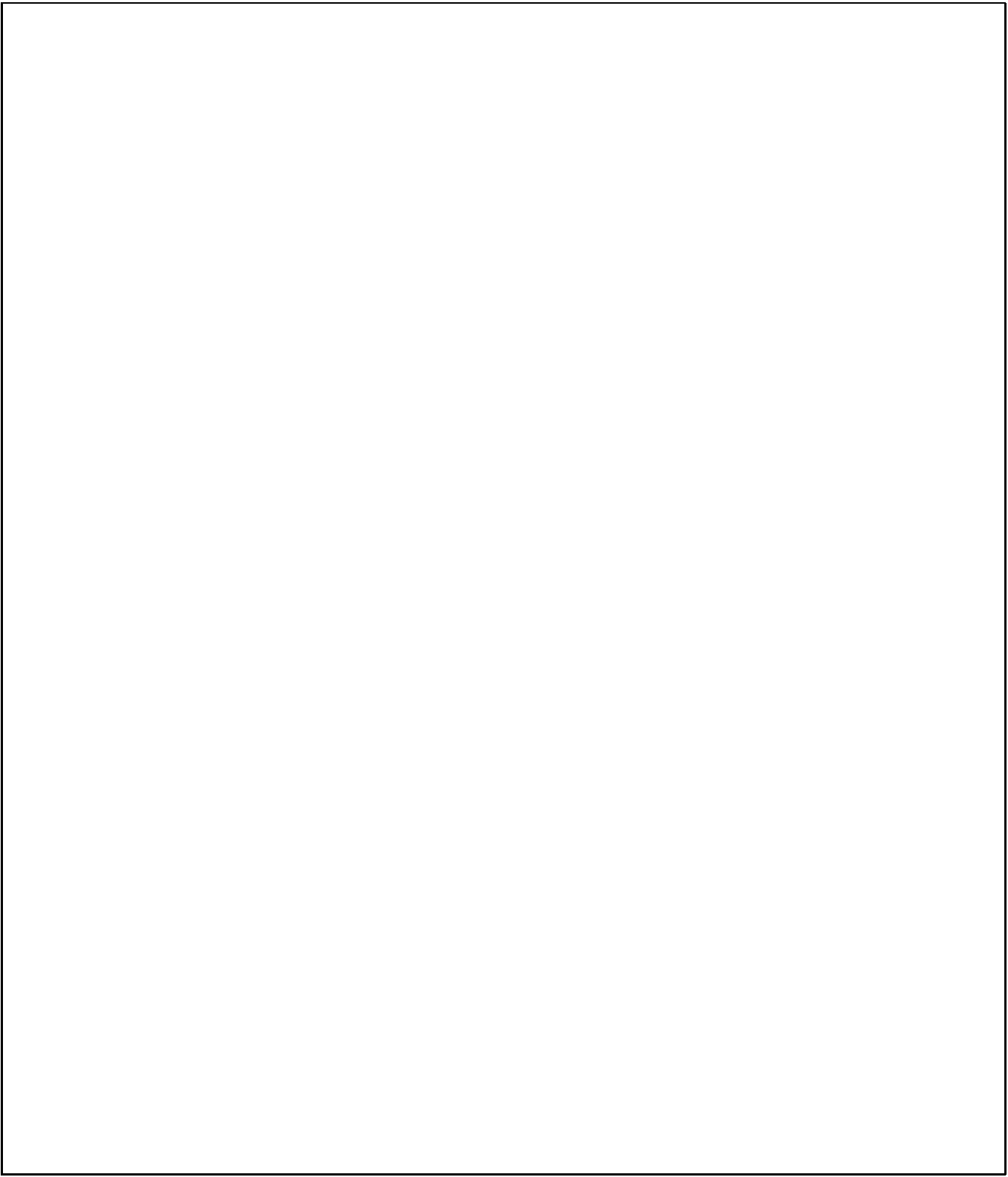

\picplace{21cm}
\caption[ ]{(a) Median filtering of the image shown in Fig. 1 with a 
box size of 9\arcs . (b) B-R colour image in a grey scale where dark is 
bluer and white is redder. Orientation as in Fig. 1.}
\end{figure*}
Hickson 96 is a hierarchical system composed of two large and two
considerably smaller members.  H96a (NGC 7674) is a luminous SBc (as
classified by Williams \& Rood 1987), while H96b (NGC 7675) is a giant E2,
and H96c is a small Sa. H96d was classified by Hickson, Kindl \& Auman
(1989) as a dwarf Im, but as discussed in \S3.5 we suggest its
reclassification as Sm.  NGC~7674 has been widely studied as a Seyfert
galaxy, and classified as Seyfert 2 (Mirabel \& Wilson 1984).  We give a
detailed study of this galaxy in Paper II where we discuss its possible
Seyfert 1 nature. 

The four galaxies show similar redshifts (see Table 2) with a mean
heliocentric velocity of 8760\kms\ and a velocity dispersion of 
$\sigma_V$ = 160\kms. 

Two long tails can be seen in Fig 1 emerging from the region between H96a
and c. One extends more than 100\arcs\ to the NE and the other is cut by
the edge of our frame at 50\arcs\ to the NW.  A shorter tail ($\sim$ 9\arcs
\, $\times$ 3\arcs ) is also present in the disk of H96a, $\sim$ 25\arcs \,
N of its center.  The beginning of a faint wide extension coming out from
H96b toward the SW is detected in all 3 filters, and it is more clearly
defined (as for the tails) after median filtering the images with a box
size of 9\arcs \, $\times$ 9\arcs . The effect of such a filtering can be
seen for the V filter in Fig. 3a. This extension was not detected in
previous observations of the group. It cannot be discarded that it extends
outside the CCD field.  There is also an indication of a bridge of matter
between galaxies a and d, although deeper images would be needed in order
to confirm it. 

Fig. 3b shows a B-R colour index image of the group in a logarithmic grey
scale representation.  Galaxies H96a and d have blue colours, while b and c
show redder ones. The measured and uncorrected colour indices of both long
tails ((B-V) = 0.5--0.6) are consistent with those of the H96a disk 
((B-V)= 0.51). 
\begin{table}
\caption[ ]{Magnitudes and colors of dwarf galaxies.}

\begin{tabular}{crcc}
\hline\noalign{\smallskip}
\multicolumn{1}{c}{Dwarf No.} & $m_B$ & (B-V)$^{(1)}$ & (V-R)$^{(1)}$ \\
\noalign{\smallskip}
\hline\noalign{\smallskip}
 1 &  20.5 &1.3 &0.8 \\
 2$^{(2)}$ &  21.1 &1.1 &0.5    \\
 3$^{(3)}$ &  21.3 &0.9 &0.7 \\
 4 &  20.5 &1.5 &1.0 \\
 5 &  20.8 &1.5 &1.0 \\
 6 &  21.2 &0.9 &0.6 \\
\noalign{\smallskip}\hline
\end{tabular}
\begin{list}{}{}
\item[$^{\rm (1)}$] Colour indexes corrected only for galactic extinction.
\item[$^{\rm (2)}$] The light of this galaxy is contaminated by H96b.
Its magnitudes were obtained after removing the r$^{1/4}$ model of H96b.
\item[$^{\rm (3)}$] This galaxy shows a two component profile.
\end{list}
\end{table}

We have identified other galaxies in the field of Hickson 96, since many
faint objects can be seen in Fig. 1.  Radial profiles could be obtained for
six of them, and allowed us to distinguish galaxies from stars by comparing
the surface brightness profile of the objects detected in the field with
that of a star (numbered as 0 in Fig. 1), which has a steeper light
distribution.  Six galaxies (numbered as 1-6 in Fig. 1a) were identified in
this way, as shown in Fig. 4. 
\begin{figure}
\epsfig{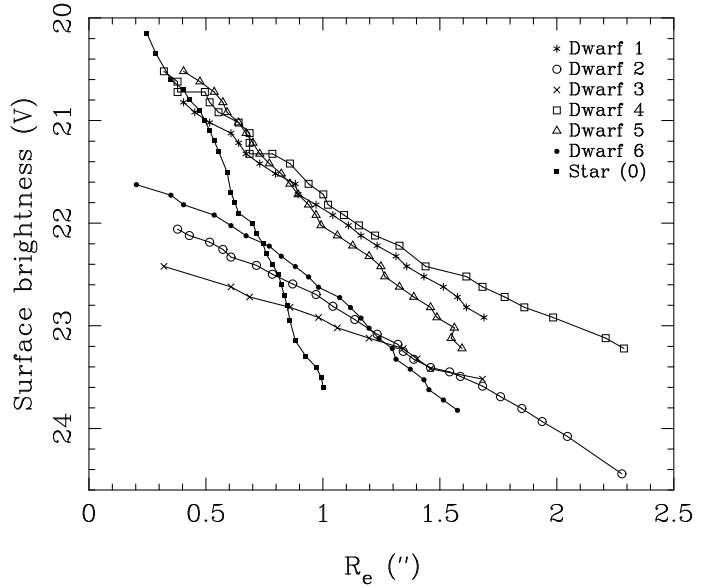}
\caption[ ]{Averaged V-band surface brightness profiles of 6 dwarf 
galaxies and a star in the frame, as a function of equivalent radius of 
each isophotal level. Dwarfs and star are located by their numbers in Fig. 1.}
\end{figure}
We give their magnitudes and colour indexes in Table 3.  No information on
their redshifts is available, so we cannot say whether or not they belong
to the group.  Inspection of their colour indices indicates that galaxies 4
and 5 are quite red and therefore are probably background galaxies
projected in the field.  Galaxies 1,2,3 and 6 are possible dwarf galaxy
members of Hickson 96.  An excess of faint galaxies compared to the
background has been reported by Carvalho et al. (1994) in Hickson groups. 
As is our case, the lack of velocity information prevents to state their
membership to the group. Hunsberger, Charlton \& Zaritsky (1996) report the
presence of dwarf galaxies in tidal tails in Hickson groups, but this is
not the case for the faint galaxies in Hickson 96. 

We have also investigated the neighborhood of Hickson 96. It is not located
in a loose group or cluster (Rood \& Struble 1994).  After inspection of
both the CfA catalog (Huchra et al. 1993) and NED\footnote[1]{The NASA/IPAC
extragalactic database (NED) is ope\-rated by the Jet Propulsion
Laboratory, California Institute of Technology, under contract with the
National Aeronautics and Space Administration} database we find that there
is no galaxy with a magnitude comparable to that of H96a and H96b within 1
Mpc in distance and 1500 \kms \, in redshift. The closest galaxy, UGC
12630, is located at 780 kpc but has a magnitude of $m_B$=15.4 mag, i.e.
1.5 mag fainter than H96a and b, and of the same order as H96c.  The next
galaxy with similar measured redshift is at a distance of 1.4 Mpc and has a
magnitude of 16.5 (the same order as H96d). The remaining 7 objects (from
NED database) within 1 Mpc do not have measured redshift and have fainter
magnitudes than the members of the group, and probably correspond to
background galaxies. 
\begin{figure}
\epsfig{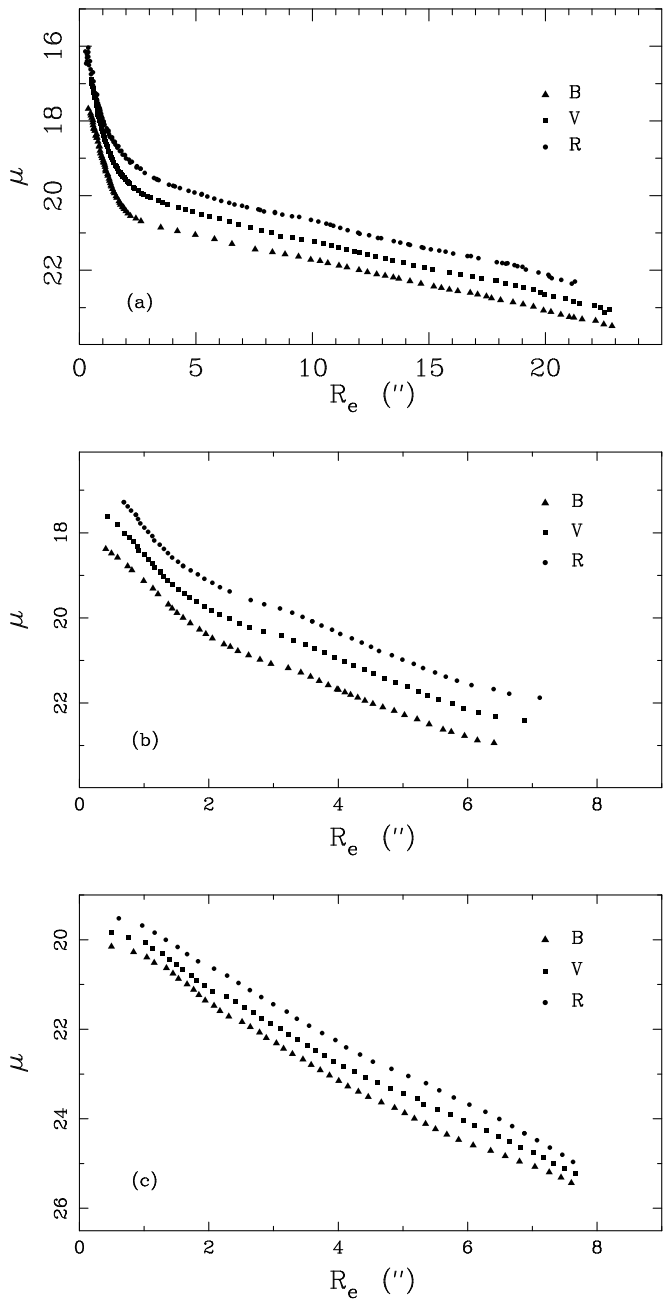}
\caption[ ]{Averaged radial profiles in each individual photometric pass 
band, as a function of the equivalent radius of each isophotal level 
for (a) H96a, (b) H96c and (c) H96d.}
\end{figure}
\begin{figure}
\epsfig{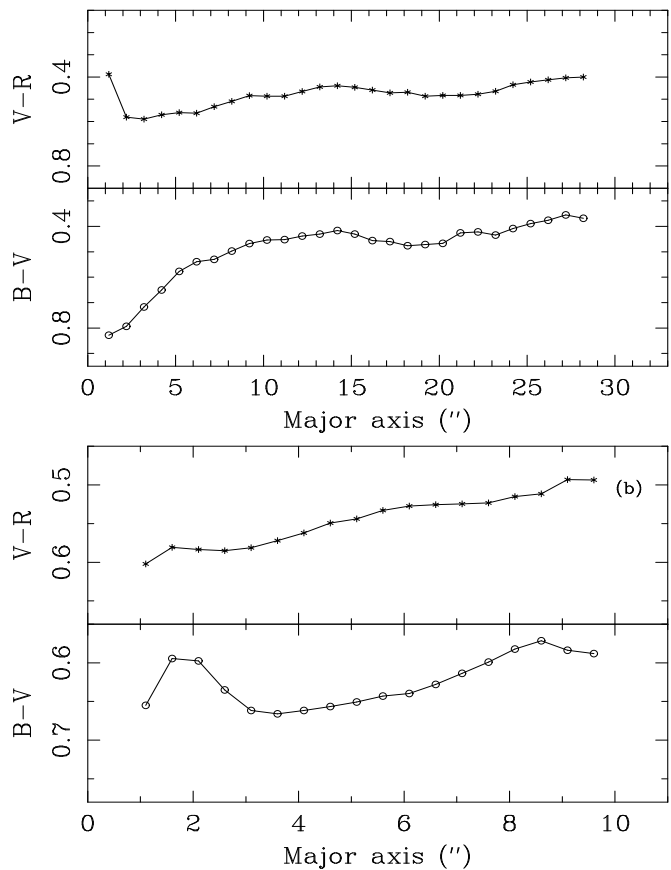}
\caption[ ]{Radial colour index profiles B-V and V-R obtained as explained 
in \S3.2. and corrected as detailed in \S2.1. (a) H96a,  (b) H96c.}
\end{figure}
\subsection{H96a}

Fig. 5a shows the radial brightness profiles for H96a in the three filters
as a function of the equivalent radius of the corresponding isophotal
level.  The signature of the Seyfert nucleus is visible at the central
parts as a pronounced steepness that adds to the bulge and disk components.
 The humps at equivalent radii of 10\arcs \, and 19\arcs \, correspond to
the spiral arms.  The disturbed morphology of this galaxy, and its strong
spiral arms prevent an accurate disk-bulge decomposition.  We have fitted
an exponential law to the disk and removed it from the profile. To the
remaining emission a r$^{1/4}$ law has been fitted. This procedure has been
performed in an iterative way until convergence is achieved.  The final
decomposition gives in B colour 
\newline
$\mu _e$ = 22.3 mag/arcsec$^{2}$, r$_e$ = 13\arcsper 7 and 
m$_d$ = 14.1 mag for the disk, and \newline
$\mu _e$ = 17.9 mag/arcsec$^{2}$, r$_e$ = 0\arcsper 5 and
m$_b$ = 16.2 mag for the bulge.

The radial colour index profiles have been calculated from the individual
B, V and R bands as follows.  The images have been integrated over circular
annuli with thickness of 1\arcs \ on the deprojected image of the galaxy
(see below for the deprojection values).  We have obtained the B-V and V-R
colour indices from these values, and corrected them as explained in
$\S$2.1. A colour gradient from redder to bluer colours with increasing
radii is seen (Fig. 6a). A colour gradient of 0.1 mg in B-V and in V-R is
found in the disk of H96a. Disk gradients have been reported for spiral
galaxies, but mostly studied in the NIR and have been assigned to 
extinction by dust in the blue bands (Peletier et al. 1994).  H96a,
although classified as unbarred in the RC3, was found to be barred by
Williams \& Rood (1987) from visual inspection of the POSS. We found for
the bar a size of 15\arcs \, $\times$ 5\arcs \, and 
a PA $\sim$ 120$^{\circ}$. 

\begin{figure}
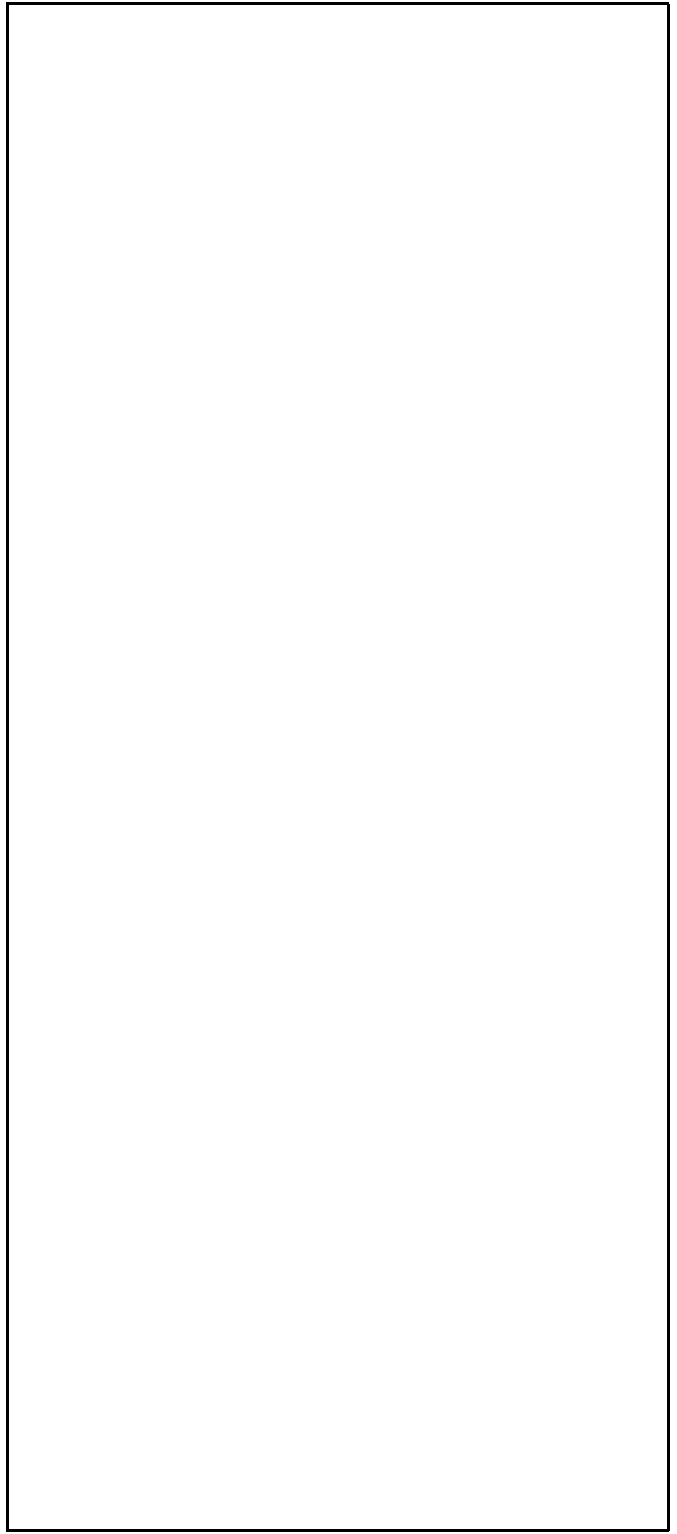

\picplace{20.3cm}
\caption[ ]{ (a) B image of H96a and c in a logarithmic grey scale 
representation. The isophote  at 26.3 mag/arcsec$^2$ is overlayed, 
together with the ellipse fitted to its non distorted portions.
(b) The same as in a). The overlapped isophotes are 20.7 and 
22 mag/arcsec$^2$. The spiral arms have been traced with dots.
(c) Sharpening of the image shown in a) obtained as explained in \S3.2. 
Darker areas correspond to excess emission.}
\end{figure}
The B band image of H96a is shown in Fig. 7a in lo\-ga\-rithmic grey scale
representation. H96c is also seen separated by 34\arcs \, from H96a.  We
note that the outer isophotes of the disk in H96a are not centered on the
nucleus but are shifted towards the side opposite H96c.  In order to
quantify this displacement we have fitted ellipses to a set of outer
isophotes.  Distorted portions of the isophotes closer to H96c have been
excluded from the fit.  One of the isophotes together with the fitted
ellipse is plotted on the image. The fit shows, for all 3 filters, a
displacement of the center of the disk of 5\arcs\ $\pm$ 0\arcsper 3 \, to
the SW, i. e.  nearly perpendicular to the bar. 
       
The outer isophotes in H96a are perturbed in the region closest to H96c. As
noted in $\S3.1$, a tail-like feature, already noticed by Arp (1966), can
be seen in the three bands. It is located 27\arcs\, north of the H96a
center. This tail has a size of $\sim$ 9\arcs \, $\times$ 3\arcs \, and is
bluer ((B-V)$_0$ = 0.1) than the disk of the galaxy ((B-V)$_0$ = 0.4). 
Opposite the nucleus and 27\arcs\, to the South, a similar, although less
blue feature ((B-V)$_0$ = 0.3), is observed.  A smaller (7\arcs \, $\times$
3\arcs) tail, also with colours similar to those of the disk, is seen
24\arcs \ to the NE of H96a. 
                                                                     
H96a has two spiral arms that become broader with increasing radius. In
Fig. 7b the arms are traced with dots, and we show also the contour up to
which the galaxy can be considered bisymmetric.  In order to enhance the
small/intermediate scale structures we have constructed a ``sharpened
image" (Fig. 7c).  We performed bayesian deconvolution with a softening
parameter $\kappa _ s = 0.025$ and subtracting one smoothed with $\kappa _
s = 0.005$, following the method developed by Molina et al. (1992).  The
bar in the center of NGC 7674 can be seen clearly in that image, as well as
the beginning of both spiral arms. Figures 7b and 7c show that the spiral
arms start from the edges of the bar, and have a m$=2$ symmetry out to r
$\sim$ 20\arcs\, with a symmetry axis in the bar direction.  At larger
radii the arm in the vicinity of H96c disappears, and that extending to the
SW of H96a is broadened. 

From our images we obtain for H96a an inclination of $\sim 31^{\circ}$ to
the line of sight and a PA of $\sim 91^{\circ}$, assuming that the outer
isophotes of the galaxy correspond to a pure disk component (Fig. 5a). The
determination of the position angle could be however inaccurate since
distortions of portions of the isophotes produced by the proximity of H96c
prevent their use in the fit.  The fact that the outer disk could be
intrinsically non-circular due to interaction suggests the use of more
internal isophotes. Those give a PA of $\sim$ 160$^{\circ}$ and an $i$
$\sim$ 44$^{\circ}$.  However the shape of the isophotes for the inner disk
is dominated by the spiral structure, and therefore should not be used for
the deprojection of the galaxy. 
\begin{figure}
\epsfig{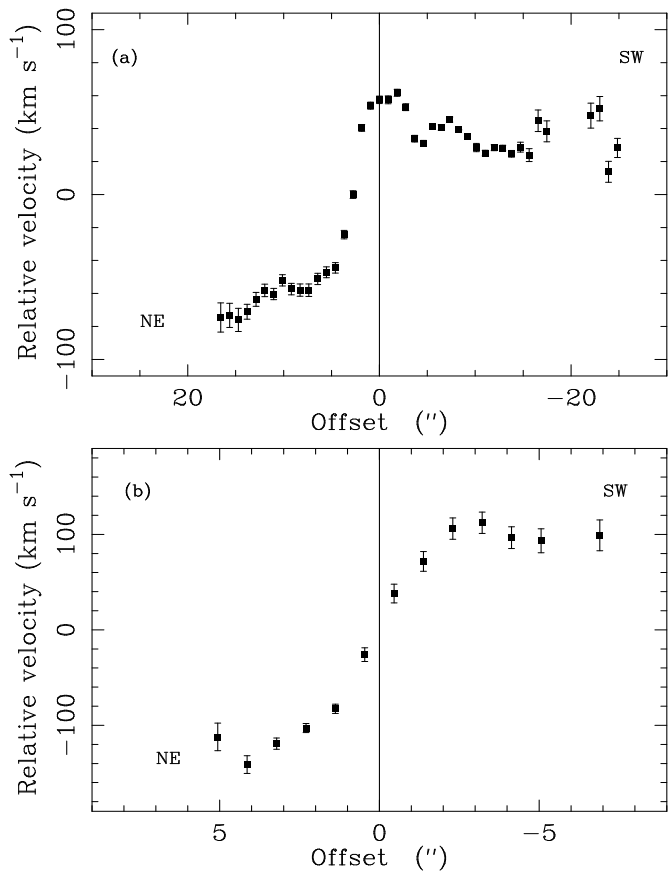}
\caption[ ]{Rotation curve of H96a (a) and H96c (b) measured along the 
line joining both galaxies, i.e. p.a. =  66$^{\circ}$.
The vertical lines  correspond to the continuum emission center.}
\end{figure}

The rotation curve in the direction joining the centers of H96a and H96c is
shown in Fig. 8.  It has been obtained from the high resolution spectrum
that contains H$\alpha$, [SII] and the [NII] lines, with the
cross-correlation technique proposed by Tonry \& Davis (1979) using the
central spectrum as template.  The 0 value in abscissa corresponds to the
continuum center position.  Fig. 8a shows that the curve extends 7\arcs \
farther in the H96c opposite direction.  We note that the kinematical
center of the curve does not coincide with the emission one, both in
H$\alpha$ and continuum, but is shifted by $\sim$ 3\arcs \, in the
direction of H96c, and has a velocity $\sim$ 30\kms \, lower. 

The gas seems to be perturbed in the outer parts, mostly toward H96c, where
it shows increasing velocities. This area corresponds to the area where one
of the arms gets disrupted.  The rotation curve also shows several minima
and maxima that correspond to the morphological structure of the galaxy. In
fact, the position of the minima at --15\arcs, --7\arcs\, and 6\arcs \,
from the kinematical center (Fig. 8a) coincide with the inter-arm regions.
The points at --28\arcs \, and --30\arcs\ from the center correspond to two
HII regions located in the spiral arm in the opposite direction of H96c. 

Using both our rotation curve (PA = 66$^{\circ}$) and those by Unger et al.
(1987; PA = 113$^{\circ}$ and 23$^{\circ}$), we find that the kinematic
axis of the galaxy must lie around a PA of 121$^{\circ}$. The best choice
for the inclination of the galaxy is $i$ = 31$^{\circ}$.  These values for
the PA and inclination are compatible with those derived from the analysis
of the outer disk isophotes. 

We have measured a semi-amplitude in the central parts of our rotation
curve of about 60 \kms . Taking into account the inclination of the galaxy,
and the fact that the slit was placed at 55$^{\circ}$ from the major axis,
we obtain a semi-amplitude after deprojection of 200\kms , a normal value
for the morphological type of this galaxy. 

We have a also carried out observations of the CO(J = 2$\rightarrow$ 1) and
(J =1 $\rightarrow$ 0) rotational transition lines toward H96a with the
30-m telescope of the Institut de Radio Astronomie Millim\'{e}trique (IRAM)
at Pico Veleta (Granada, Spain). We have obtained a total H$_2$ mass of
2$\times$10$^{10}$ M$_{\odot}$ assuming a H$_2$ to integrated CO flux ratio
of 3.6$\times$10$^{20}$ cm$^{-2}$(K \kms )$^{-1}$ (Dickman et al. 1986). 
Both CO maps show enhanced emission that seems to be associated with the
spiral arms, and the gas follows the overall rotation of the galaxy.  A
detailed description of these observations and results are given in a
following contribution (hereafter Paper II), since the coverage includes
only the central part of H96a. 

\subsection{H96c}

In the image of the galaxy shown in Fig. 7b and c it can be seen that H96c
shows a disk-bulge morphology with a two-armed spiral at the inner parts. 
One of the arms however disappears at $\sim$ 2\arcsper 5 from the center of
the galaxy in the direction of H96a. A bar could also be present, as
suggested by the elongated structure seen in the sharpened image shown in
Fig. 7c, but the high inclination of the galaxy together with its small
size do not allow a sure identification. 
 
The surface brightness profiles of H96c in the three filters are shown in
Fig. 5b.  The galaxy shows an excess of light produced by its spiral
structure in an intermediate region (2\arcs \, \< r$_e$ \< 4\arcs ) and,
although the bulge and disk can be clearly noticed in the profile, a
reliable quantitative decomposition is not possible due its high
inclination. Colour index profiles (Fig. 6b) have been obtained as for
H96a, with the deprojection parameters given below. H96c shows a rather
steep colour index gradient, becoming bluer at larger radii.  In addition
it has perturbed outer isophotes and two bluer protuberances are noted
towards the direction opposite to H96a. Their colour indices are slightly
bluer ((B-V)$_0 = 0.5$, (V-R)$_0 = 0.4$) than those of the disk 
((B-V)$_0 = 0.7$, (V-R)$_0 = 0.6$). 

Both the protuberances and the spiral structure of H96c make it difficult
to determine accurately the isophote centers. However we have found no
important isophotal off-centering either before the beginning of the spiral
arms at re $\sim$ 1\arcsper 7, or for the outer isophotes free from
perturbations (4\arcs \< r$_e$ \< 6\arcs ).  For larger radii the proximity
of H96a prevents a fit.  The ellipticity is constant in the central parts,
with a value of 0.25, increasing to 0.4 in the outer parts, where only the
disk is present. This shows the existence of an important bulge in the
inner parts as expected for an early-type spiral and a possible hint of the
presence of a central bar.  We have used the outer unperturbed isophotes
for the calculation of the deprojection parameters of H96c, obtaining an
inclination of 52$^{\circ}$ and a PA of 33$^{\circ}$. 
\begin{figure}
\epsfig{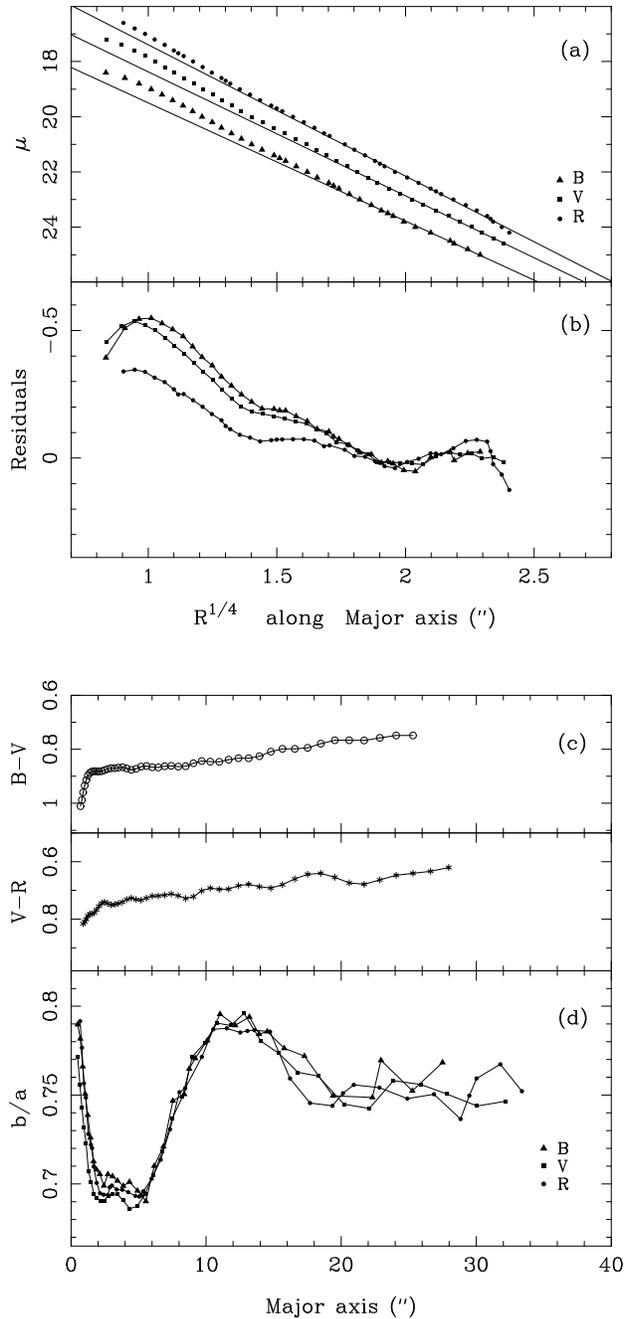}
\caption[ ]{ (a) Surface brightness profiles of H96b in each filter as 
a function of the major axis radius of each isophotal level. We have  
superposed the r$^{1/4}$ laws fitted to these profiles. (b) Residuals
from these fits, i.e. the  difference between the observed radial profile 
and the fitted one. (c) Radial colour index profiles (B-V) and (V-R) 
obtained as explained in \S3.2. and \S2.1. (d) Variation of the axial 
ratio of the ellipses fitted to the isophotes as a function of their 
semimajor axis.}
\end{figure}

The spectrum of the galaxy shows prominent absorption bands and lines,
typical of an old stellar population with a large break at 4000\AA.
Following the indices by Pickles (1985) the galaxy has a spectral type
between G0 and G8 and over-solar metallicity.  Overlapped to the old
stellar population the galaxy shows emission lines indicating that a burst
of star formation is occurring as found by Laurikainen \& Moles (1988). 
The emission lines like H$\alpha$ and [NII] extend for about 6 kpc. In Fig.
8b we show the rotation curve obtained with the high resolution spectrum by
using H$\alpha$, [NII] and [SII] lines.  The direction of the slit was that
of the line connecting the centers of H96a and H96c. The kinematical and
geometrical centers are basically coincident within our spatial resolution.
 In fact, as can be seen when we symmetrize the curve, the section of
maximum emission is located at 0\arcsper 5 from the kinematical center. 
The curve is symmetric in the central parts, but some perturbations seem to
occur for the outer ones.  The relative velocities remain constant toward
H96a, while they might decrease toward the opposite direction.  The change
in velocity, however, is within the error bars.  The deprojected
semi-amplitude of the velocity curve is 317\kms . 
         
\subsection{H96b}

The galaxy H96b (NGC~7675), an early type object, is the second brightest
galaxy in the group, and its center is located {2\arcmper 3} to the SE away
from the center of H96a. In Fig. 9a we show the azimuthally averaged
surface brightness profiles obtained for the three photometric filters. 
The profiles of the outer regions are well fitted by a \rdv\ law. However
in the central region, the galaxy exhibits sinusoidal deviation from the de
Vaucouleurs profile with amplitude increasing towards the center.  This is
illustrated in Fig. 9b where we plot the difference between the observed
brightness profile and the \rdv\ law obtained when fitting for distances
larger than 10\arcs.  Different effective radii have been obtained for each
filter, 
\newline 
$r_e = 9\arcsper 2\pm0\arcsper 6$ and $\mu _e$=$23.1\pm0.2$ for B filter, 
\newline
$r_e = 7\arcsper 1\pm0\arcsper 3$ and $\mu _e$=21.6$\pm$0.1 for V filter, and 
\newline
$r_e = 6\arcsper 0\pm0\arcsper 2$ and $\mu _e$=20.5$\pm$0.1 for R filter. 

The colour index profiles indicate that NCG 7675 becomes increasingly
redder towards the center (Fig. 9c).  The colour index images show a red
structure that extends approximately along the major axis.  (light central
region in Fig. 3b) with an axial ratio of about 0.4.  It could indicate 
the presence of dust. 
\begin{figure*}
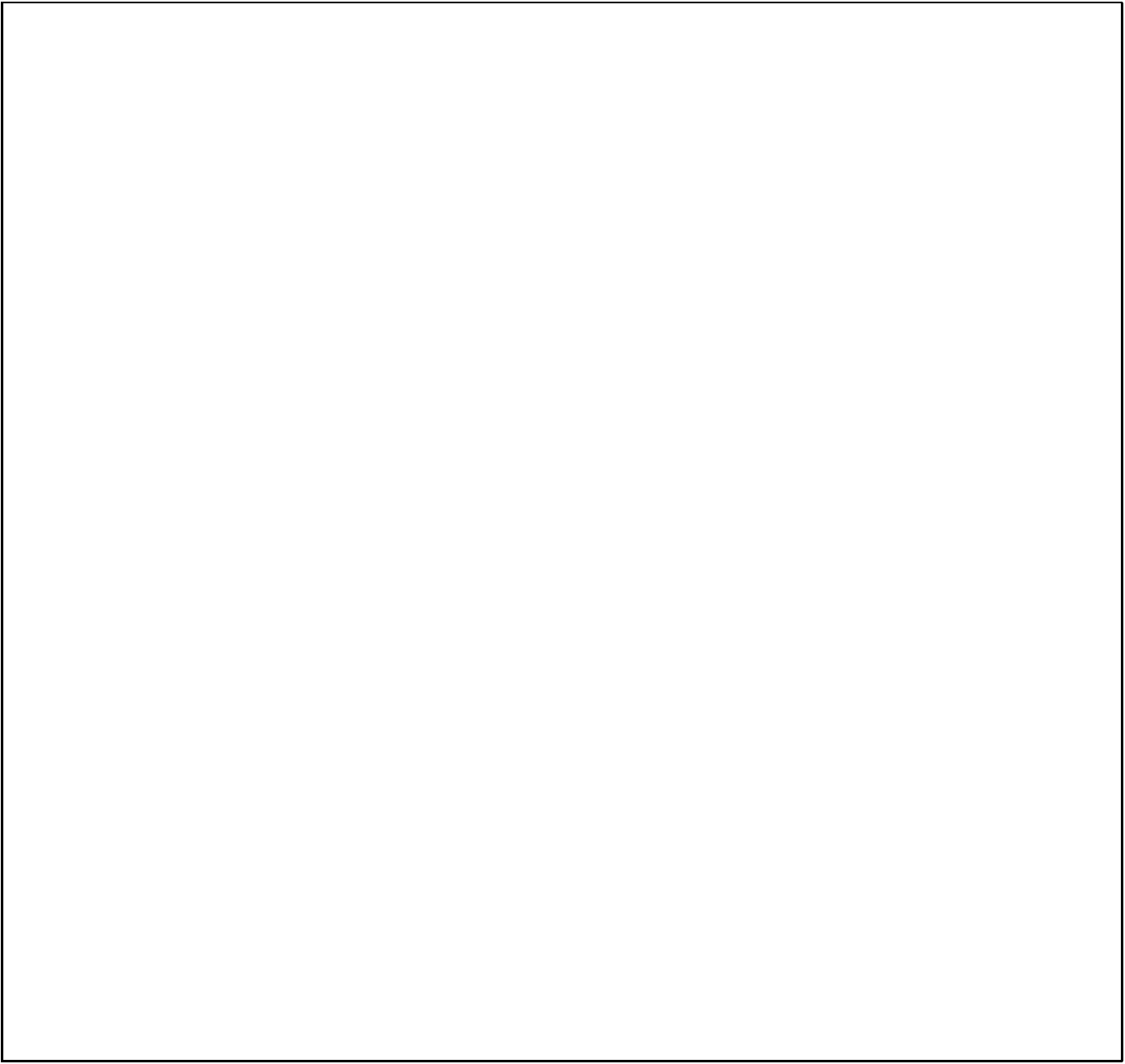

\picplace{17.cm}
\caption[ ]{(a) Isophotal contours corresponding to the V image 
of NGC 7675 for the inner 30\arcs. (b) Residuals obtained by the subtraction 
of a 3\arcs$\times$3\arcs\, box median filtered image from the V image.
(c) Isophotal contours corresponding to the $r^{1/4}$ profile best fitting 
the main body of the galaxy convolved with the seeing. (d) Residuals 
obtained by the  subtraction of the fitted model in (c) from the  V image 
in (a). Dwarf 2 is clearly seen after subtraction of the main body of H96b. }
\end{figure*}

We performed a quantitative analysis of the shape of the galaxy by least
squares fitting of a set of isophotes to ellipses following the well known
method proposed by Carter (1978) and widely applied by Bender et al.
(1989). The results are displayed in Fig. 9d, where we plot the axial ratio
as a function of the semimajor axis.  Excluding radii smaller than
1\arcsper 2, where the ellipticity is dominated by the seeing, there is a
region between $r_e$ = {1\arcsper 2} and {6\arcs} where $\epsilon$ is
nearly constant in all three filters with a value of $\epsilon = 0.30$. 
This region has a larger ellipticity than the rest of the galaxy, and is
also the same region where the residuals with respect to the \rdv\ become
more significant.  This is opposite to the seeing effect, which makes
constant ellipticity isophotes to appear rounder toward the center.  Then
an abrupt variation to rounder isophotes ($\epsilon = 0.20$) is seen from
6\arcs \, to 12\arcs , to reach an almost constant value of $\epsilon =
0.25$ at about 16\arcs\ from the center. 

All the isophotes are concentric for all filters.  Their position angle is
nearly constant at PA$= 52\deg$ for the inner 20\arcs\ and decreases to
\approx \ 46\deg at 30\arcs. Its determination, however, becomes less
accurate with increasing radius since the ellipticity is lower in the outer
regions. The same applies when determining the fourth coefficient of the
cosine term, A$_4$, in the Fourier analysis of the residuals with respect
to ellipses. Caon et al.  (1990) determined that $\epsilon \ge 0.3$ is
necessary in order to obtain significant values for A$_4$.  In our case it
was only possible to determine A$_4$ out to 15\arcs\ from the center.  We
do not detect any significant trend until 5\arcs, and only in the region
where $\epsilon$ changes rapidly is it noted that A$_4$ becomes positive.
For larger distances, however, this coefficient is not statistically
significant in the Fourier expansion. 

We have further analyzed the morphology of the galaxy (Fig. 10a) in two
other ways. The first method was to smooth the image by filtering by a
running median box of 3\arcs\ of side, and to subtract the result from the
original image (Fig. 10b). The second approach was to subtract a model of a
seeing convolved \rdv\ galaxy (Fig. 10c) with the parameters obtained from
the external isophotes of the original galaxy, i.e. $\epsilon = 0.25$ and
PA$= 52\deg$. The subtraction is shown in Fig. 10d. In both cases there
exist significant residuals in the central part of the galaxy, where we see
a structure of \approx \, 7\arcsper 5$\times$3\arcs\ in size plus some
residuals for the central 10\arcs, while the outer part of the galaxy
vanishes, since it is well represented by a \rdv\ law. This inner feature
is detected in the three filters. 

A faint plume also appears when we analyze the median smoothed V image that
is displayed in Fig. 3.  It is also seen in the B and R filters. Another
deep image of the group, but with a larger field of view, would be needed
to characterize its properties, in particular its extension and colour
indexes.  A second tidal feature, a bridge of optical light, is also
visible in the same image, joining H96b to H96d. It is also detected in the
B image, and more strongly in the R image.

The spectra of the galaxy show only absorption lines with no signature of
recent star formation.  The velocities and velocity dispersion along the
major and minor axes of H96b, were obtained by the Tonry \& Davies (1979)
method. The velocity, velocity dispersion and their errors were determined
by cross--correlation of the spectra in each spatial section with those of
15 template stars.  The final values for each spatial section were then
obtained by weighting by the errors. In Fig. 11a and 11b we present the
velocity curve along the slit for the major and minor axes respectively.
For the minor axis, the radial scale was divided by the axial ratio in
order to compare directly with the curve obtained along the major axis. The
error bars in the figure correspond to the mean quadratic error for each
point. As can be seen from the figure a rotation is observed in the major
axis for the central 10\arcs\ where the S/N is high, and it amounts to
\approx \, 30-40\kms.  The velocity curve is perturbed in the minor axis
direction, being decoupled from the rest of the galaxy for the region
within 6\arcs \ from the center. 
\begin{figure}
\epsfig{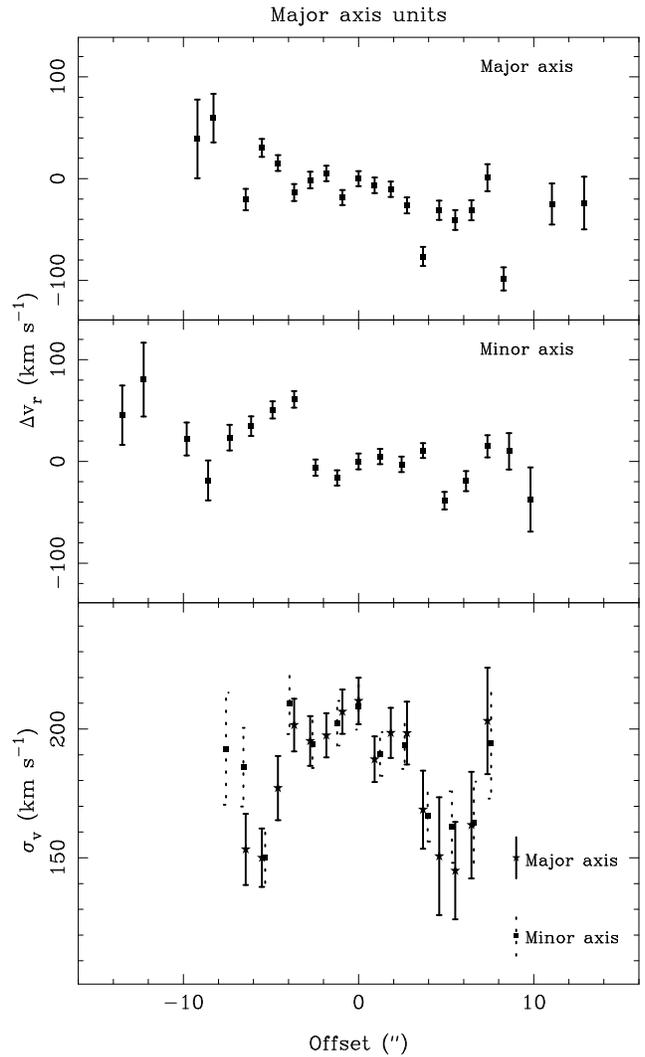}
\caption[ ]{Velocities along the major axis (a) and the minor axis 
(b) of H96b as a function of the distance to the center in major axis units. 
(c) Velocity dispersion profiles along major axis (stars and full lines for 
the error bars) and minor axis (filled squares dashed lines) of H96b.}
\end{figure}

The velocity dispersion profiles along the principal axes are shown in Fig.
11c, where the distance scale along the minor axis has been corrected as
before.  The agreement between the two axes has to be emphasized. In the
central region the velocity dispersion may be considered roughly constant
with a value of $\sigma_v \sim 200\kms$, but some structure appears. Then
the velocity dispersion falls to about $\sim 150\kms$ at a radius of
4\arcs\ and seems to increase again in the outer parts starting from a
radius of 6\arcs\ from the center. The global behavior of the velocity
dispersion profile, as well as the structure seen at the inner parts, seem
to be real since they appear in the two axes and in spectra which were
obtained in different conditions. 

\subsection{H96d}

This galaxy, the bluest of all in the group ((B-V)$_0$ = 0.17, (V-R)$_0$ =
0.40), was previously classified as an irregular Im because only the knotty
inner part was detected.  However the outer isophotes that we detect here
look quite symmetric and in fact the data are consistent with a late type
spiral with an exponential disk. In Fig. 5c we present the surface
brightness profiles in the three filters, and as it can clearly be seen
from the figure, they agree well with an exponential disk with some
perturbations due to the knots which are observable on the images. 
Assuming that the external isophotes correspond to an intrinsically
circular disk we obtain for the position angle and inclination respectively
10$^{\circ}$ and 54$^{\circ}$. 
\begin{figure*}
\epsfig{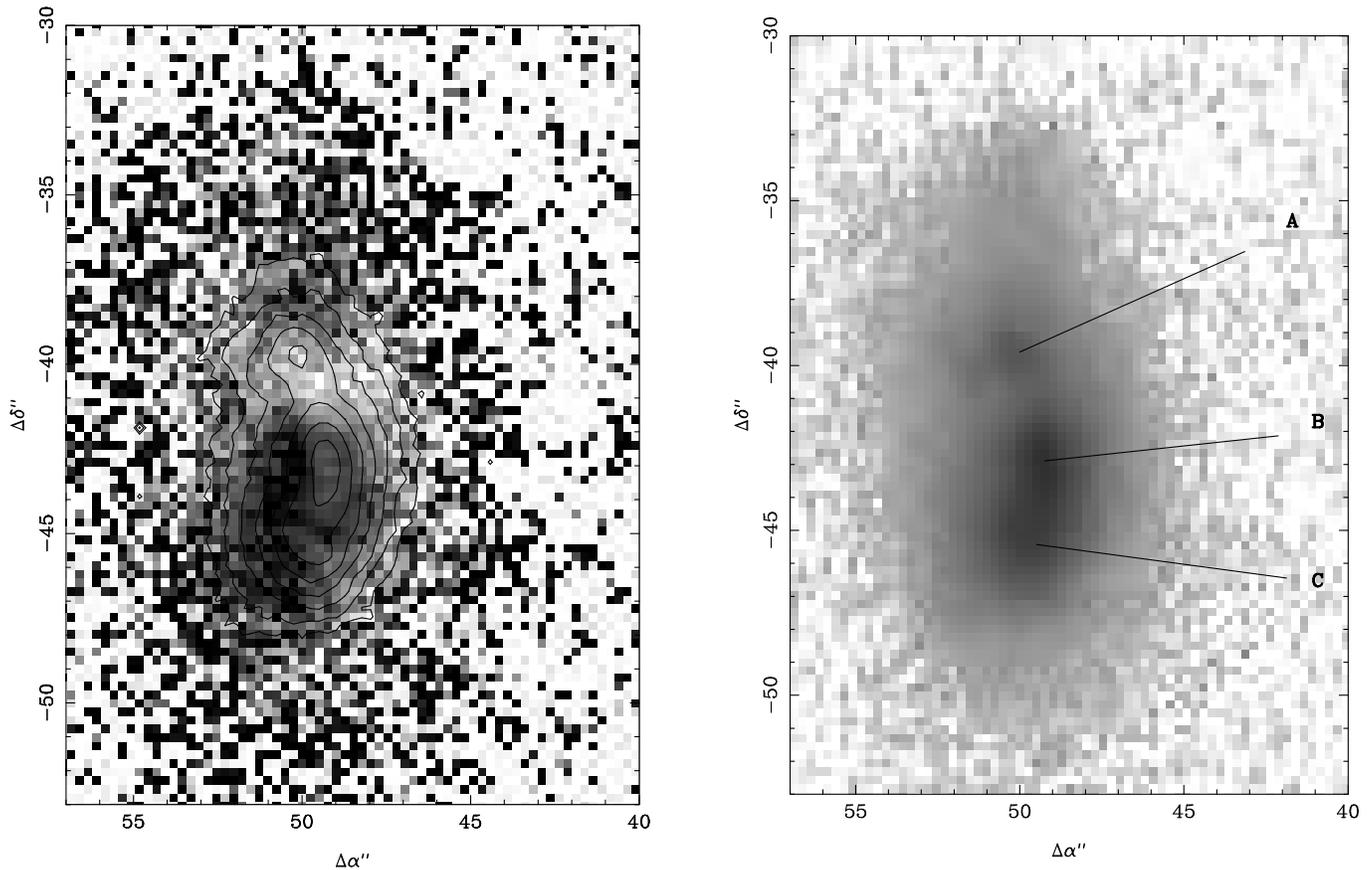}
\caption[ ]{(a) B-R colour index image of H96d in a grey scale where black
is bluer and white is redder. We have superposed R band isophotes ranging
from 16.5 to 19.5  mag arcsec$^{-2}$ with a step of 0.5 mag/arcsec$^2$. (b)
V band image of H96d where the knots referenced as A, B and C in the text
are marked.} 
\end{figure*}

In Fig. 12a we plot the (B-R) colour index image of the galaxy, together
with isophotal contours in the R filter, while in Fig. 12b we show its 
V image. 
\begin{table}
\caption[ ]{Line parameters for H96d.}
\begin{tabular}{lcrr}
\hline\noalign{\smallskip}
Line& $\lambda _0$ &  F$_{obs}$ & I$_{dered}$\\
\noalign{\smallskip}
\hline\noalign{\smallskip}
$[$OII$]$&     3727.30& 571.2& 635.5\\
H$\gamma$& 4340.50& 34.2& 36.1\\
H$\beta$&  4861.30& 100.0& 100.0\\
$[$OIII$]$&    5006.80& 142.3& 142.3\\
H$\alpha$& 6562.80& 281.0& 281.0\\
$[$NII$]$&     6583.60& 84.2& 84.2\\
$[$SII$]$&     6724.00& 89.3& 89.3\\
\noalign{\smallskip}
\hline
\end{tabular}
\end{table}
The innermost structure of H96d as seen in the three filters consists of 3
knots (named A, B and C in Fig. 12b), which are nearly aligned with the
galaxy major axis. Knots A and B seem to be separated by a dust lane (Fig.
12a), seen as a lack of light in B, V and R, and with red colour indexes
((B-V)$_0$ = 0.38, (V-R)$_0$ = 0.53). Knot A shows redder colour indexes
((B-V)$_0$ = 0.34, (V-R)$_0$ = 0.50) than B and C, which have similar
values ((B-V)$_0$ = 0.18, (V-R)$_0$ = 0.34). The eastern edge of knots B
and C is surrounded by a bluer area ((B-V)$_0$ = 0.00, (V-R)$_0$ = 0.38). 
All of this indicates a very blue galaxy with a population of young stars.
The redder colours of knots B and C and even more so of knot A are the
consequence of the presence of very intense emission lines that contribute
to the red part of the spectrum, reflecting a very recent burst of star
formation. 

For this galaxy we have two low resolution spectra at PA = 88$^{\circ}$ and
161$^{\circ}$. In the first case the slit passes through knots B and C, but
it was taken under bad weather conditions. For the second, the slit goes
through the disk and covers basically the B and C knots and part of A. 
Since it was taken under not very good seeing it was not possible to
spatially resolve the different regions. To derive physical parameters we
added all the spatial sections along the slit. In this case we note the
presence of a faint stellar continuum (Fig. 2d). The spectrum also shows a
depression to the red of H$\gamma$ due to a weak G--band. Since the
4000\AA\ break is not clearly visible, the underlying dominant stellar
population is earlier than G.  This result is also supported by the
presence of absorption for the hydrogen lines until H$\gamma$. 

Over the stellar continuum we see emission lines of [OII], [OIII], [NII],
[SII] and the Balmer lines H$\gamma$, H$\beta$ and H$\alpha$. In Table 4 we
give the observed and de-reddened intensities of the lines.  The values and
the non existence of the [OI]6300\AA\ line show that the gas is emitted by
classical HII regions with a thermal origin.  The value of 1.34 for the
ratio of the [SII] lines corresponds to a density of 100 cm$^{-3}$. By
using the empirical relations by Pagel et al. (1979) and the curves by
Edmunds \& Pagel (1984) we found T$_e$ \approx \, 10500 $\pm$ 1500K and an
abundance 1/3 solar.  From these values we have calculated the absolute
flux of the H$\beta$ continuum I$_{cont}$ = 5.62$\times$10$^{-16}$
ergs$^{-1}$cm$^{-2}$A$^{-1}$.  For the adopted distance the total flux
emitted by both knots B and C is 3.6$\times$10$^{39}$ergs$^{-1}$ and the
Lyman photon flux is 6.7$\times$ 10$^{52}$phs$^{-1}$. Given the low value
for the equivalent width for H$\beta$, the observed star formation should
have happened in a burst about 8 $\times$ 10$^{6}$ yrs ago.  The effective
temperature for the stars of the burst is $\sim$ 30000 K, corresponding to
the T$_{eff}$ of B0.  These results indicate an important stellar formation
in H96d.  The condensations may represent different bursts of induced star
formation that took place in this dwarf galaxy. 

\section{Discussion and Conclusions}
\subsection{Individual members of Hickson 96}

We have shown that the geometrical center of the H96a disk is shifted by
5\arcs\ relative to the photometric center, in the direction opposite to
H96c and perpendicular to the observed bar.  One of the two arms of H96a
vanishes in its outer parts, and the same occurs for H96c, both in the 
area where their disks overlap. An offset of the H96c disk would be 
difficult to detect due to the small size of this galaxy. Galaxies whose 
centers of the disk are offset from the center of the bar are not a rare 
case, and the asymmetry is more important for the later Hubble types, being 
often associated to one-armed spiral structures (see review by Odewah, 1996). 
Their origin is not well understood. Impact with a companion can produce
them, as shown by Athanassoula (1996) with the help of numerical 
simulations, but other mechanism should exist since isolated asymmetric
galaxies have been found (e.g. Phookun et al. 1993).  In the case of 
H96a and c, the interaction hypothesis seems plausible, since both 
galaxies are bi-symmetrical in their inner parts, while they loose their 
symmetry in the area where their disks are in contact. It is also there 
where the tails start. 

H96b could, at a first glance, appear unaffected by the other members of
the group.  We have, however, shown that this elliptical galaxy has clear
morphological and ki\-ne\-ma\-ti\-cal perturbations.  Its surface
brightness profile, while well represented by a r$^{1/4}$ law in the outer
parts, deviates from it in a sinusoidal way with the highest excess of
light in the inner 10\arcs\ ($\sim$ 4 kpc).  For the same inner radii the
ellipticity of the isophotes suffers an important increase relative to the
outer parts.  These results, together with the kinematical perturbations
found for the same range of radii, clearly suggest the presence of a very
elongated component in the center of H96b, kinematically decoupled relative
to the overall velocity pattern of the galaxy. 

These properties have been observed in other elliptical galaxies. Nieto et
al. (1991) found that all ellipticals in his sample with decoupled cores
(33\%) show photometric structures in a similar range of radii.  Forbes et
al. (1995) suggest from their observations that kinematically-distinct
cores show disk-like shapes, although the presence of dust and/or low
inclinations of the disk can make the detection of those disks difficult
(Forbes et al. 1995; Kormendy et al.  1994). This might be the case for
H96b, for which we find indications of a small--scale elongated dust lane
towards the center ($\S$3.4.). In fact higher ellipticities of the central
isophotes, as we have measured in H96b, have been attributed in the
literature to evidences of disks (Franx \& Illingworth, 1988; Forbes 1994;
Forbes et al. 1995) finding in some cases bar-like structures (Surma \&
Bender 1995). 

Several models have been proposed to explain the origin of these kind of
features, and we examine here whether they may apply to H96b. Kormendy
(1984) and more recently Balcells \& Quinn (1990) explain decoupled cores
by the capture of a small spherical/compact galaxy. Since H96b is found in
a group with likely existence of dwarf galaxies, it seems reasonable that
this giant elliptical had previously swallowed a smaller companion causing
the observed kinematical perturbations, and the observed wave--like
perturbation of its light profile.  The recent model by Hau \& Thomson
(1994) explains these signatures through a retrograde interaction with a
secondary galaxy without invoking a merger.  A completely different
approach is the streaming motion model by Statler (1991) which does not
consider an external origin.  In the case of H96b the merger/interaction
models are also favoured by its broad and diffuse tidal-like structure,
that has been predicted by simulations as an interaction signature (e.g.
Barnes \& Hernquist 1992 and references therein) and photometrically
described by Schombert et al. (1990). 

It is difficult to establish the existence of peculiarities in H96d due to
its small size, that usually implies a barely defined shape. As shown in
$\S$3.5, it is a disk system, pro\-bably a Sm galaxy, showing knots related
to the presence of bursts of star formation. This young population still
keeping ionized gas is mainly distributed in the side closer to H96a, and
might be therefore induced by this galaxy.  In fact, the knotty aspect of
H96d is quite similar to that of IC 5283, for which the existence of local
episodes of star formation processes, induced by the interaction with the
companion galaxy NGC 7469, has been claimed to explain its morphology
(M\'{a}rquez \& Moles 1994). 

\subsection{Intragroup medium and dynamical state of the group}

The morphological and kinematical perturbations of the galaxies in Hickson
groups strengthen the argument that they are real entities, and help to
define their dynamical state. Hickson (1990) estimates that one third of
the galaxies in Hickson groups show morphological disturbances, and one
third of all groups contain at least 3 morphologically disturbed
ga\-la\-xies. The ratio of perturbed galaxies is larger (2/3) on the basis
of spectroscopic data (Rubin, Hunter \& Ford 1991).  Tikhonov (1990)
suggests that 53\% of the groups he observed show clear signs of
interactions, but the majority being in pairs (64\%).  Deeper observations
of compact groups however should change these ratios.  Hickson 96 is an
example of a group with an evident interacting pair of galaxies but, as we
conclude from our observations, not only the pair but all of its members
are suffering the effects of gravitational interaction. 

Toomre \& Toomre (1972) showed that tails, bridges or plumes are signs of
gravitational interaction.  In Hickson 96 two long tails emerge from the
region between the spiral galaxies a and c, while a plume comes out from
the elliptical b which in turn is connected with the dwarf disk galaxy d
through a bridge of matter (Fig. 3).  The existence of these features
argues very strongly for the fact that Hickson 96 is a physical interacting
group.  Those features fit well with the extensive photometric study of
tidal features in interacting galaxies performed by Schombert et al.
(1990), which indicates that the internal velocity dispersion of the galaxy
from where the material originates strongly influences the appearance of
the tidal feature. Sharp features are more abundant in spiral systems,
while broad, diffuse features are more likely associated to hot components,
as is here the case. 

A diffuse intragroup component has been detected at different wavelengths
toward several Hickson groups (see $\S$1).  We have not detected diffuse
light toward Hickson 96 in the B, V and R bands at 3$\sigma$ levels of 27.2
mag arcsec$^{-2}$, 27.1 mag arcsec$^{-2}$, and 26.9 mag arcsec$^{-2}$
respectively.  Neither has diffuse X-ray emission associated to the group
been reported, although it has been observed with the ROSAT Position
Sensitive Proportional Counter (PSPC) with an exposure time of 3800s. 
This, however, cannot exclude the existence of such emission since
simulations made by Diaferio et al. (1995) indicate that a time longer than
10$^4$s would be necessary in order to detect hot intragroup gas.  X-ray
emission has been only detected toward H96a (NGC 7674), probably originated
from its Seyfert nucleus, similarly to X-ray detections of active galaxies
in H4, H16, H44 and H91 (Ebeling et al. 1994, Saracco \& Ciliegi 1995,
Pildis et al. 1995a).  Dynamical evolution would be however expected to
occur, since the material being ejected from the individual galaxies as
tidal features into the intragroup medium could build a diffuse light
component in the future evolution of Hickson 96. 

The observed harmonic radius amounts to $R_H=28.3$ kpc. This value,
together with the velocity dispersion obtained in $\S$3.1, implies a virial
mass of $M_V$ = 1.7$\times$ 10$^{12}$\solmass\ and a crossing time of
$t_{cr} = 0.03$H$_0^{-1}$. The group luminosity, obtained as the sum of the
light of the individual galaxies extrapolated to infinite, is L$_B$ \approx
7 $\times$ 10$^{10}$\solum \ which correspond to the rather low value for
the mass luminosity ratio of 20\solmass/\solum.  The low velocity
dispersion and crossing time contradicts in the case of Hickson 96 the
argument that groups are chance superposition of pairs and non related
galaxies within loose groups. 

From the point of view of isolation Hickson 96 is a well isolated group
since it is not part of any loose group or cluster and only two galaxies
fainter than H96a and b can be found in a radius of 1 Mpc and with a
difference in velocity smaller than 1500 \kms \ ($\S$3.1). 

\subsection{Concluding remarks}

We present here evidence that Hickson 96 is a dynamical entity.  Tidal
features associated to all of its members, a small velocity dispersion
(160\kms), together with its high degree of isolation argue very strongly
against alternatives for the origin of Hickson 96 such as a transient group
in a loose one or chance projections.  Therefore we conclude that the most
probable alternative is that Hickson 96 is a real compact group, where the
observed features can be accounted for by multiple tidal interactions in a
dense environment such as that of compact groups. 
       
\begin{acknowledgements}                  
We acknowledge helpful discussions with Dr. J. Sulentic.  This work was
partially supported by DGICYT Spanish Grant PB93-0159 and by Junta de
Andaluc\'{\i}a (Spain). 
\end{acknowledgements}


\begin{thebibliography}{}

\bibitem{} Arp, H. C. 1966, in Atlas of Peculiar Galaxies (California 
Institute of Technology, Pasadena)
\bibitem{} Athanassoula, E.  1996, in ``Barred Galaxies", eds. R. Buta, 
D.A. Crocker \& B.G. Elmegreen, ASP Conference Series, Vol. 91, 30
\bibitem{} Athanassoula, E., Makino, J. \& Bosma, A.  1996, preprint
\bibitem{} Bahcall,  N. A., Harris, D. E. \& Rood, H.J. 1984,  ApJ 284, L29
\bibitem{} Balcells, M. \& Quinn P.J., 1990 ApJ. 361, 381
\bibitem{} Barnes, J. E. \& Hernquist, L.  1992, ARA\&A 705, 1
\bibitem{} Bender, R., Surma, P., Doebereiner, S.,  Moellenhoff, C. 
\& Madejsky, R. 1989, A\&A 217, 35
\bibitem{} Burstein, D. \& Heiles, C. 1984, ApJS, 54, 33
\bibitem{} Caon, N., Capaccioli, M. \& Rampazzo, R.    1990, A\&AS 86, 429
\bibitem{} Carter, D.  1978, MNRAS 182, 797
\bibitem{} de Carvalho, R. R., Ribeiro, A. L. B. \& Zepf, S. E. 
1994 ApJS 93, 47
\bibitem{} Dell'Antonio, I. P., Geller, M. J. \&  Fabricant, 
D. 1995, AJ 110, 502
\bibitem{} Del Olmo, A., Moles, M. \& Perea, J. 1995, in ``Groups of 
Galaxies", eds. O.G. Richter \& K. Borne. ASP Conf.Series Vol. 70, p. 117
\bibitem{} de Vaucouleurs, G., de Vaucouleurs, A., Corwin, H. G., Buta, R. J.,
Paturel \& G.,  Fouqu\'{e}, P. 1991, Third Reference Catalogue of Bright
Galaxies. (New York: Springer--Verlag) (RC3)
\bibitem{} Diaferio, A., Geller, M.J. \& Ramella, M. 1994, AJ 107, 868
\bibitem{} Diaferio, A., Geller, M. J. \& Ramella, M.  1995, AJ 109, 2293
\bibitem{} Dickman, R. L., Snell, R. L. \& Scholerb, F. P. 1986, ApJ 309, 326
\bibitem{} Ebeling, H., Voges, W. \& B\"ohringer, H. 1994, ApJ 436, 44
\bibitem{} Edmunds, M. G. \& Pagel, B. E. J.  1984, MNRAS 211, 507
\bibitem{} Forbes, D. A., Franx, M. \& Illingworth, G. D.  1995, AJ 109, 1988
\bibitem{} Forbes, D. A.  1994, AJ 107, 2017
\bibitem{} Franx, M. \& Illingworth, G. D. 1988, ApJ 327, L55
\bibitem{} Hau, G. K. T. \& Thomson, R. C. 1994, MNRAS 270, L23
\bibitem{} Hernquist, L., Katz, N. \& Weinberg, D. H. 1995, ApJ 442, 57
\bibitem{} Hickson, P. 1982, ApJ 255, 382
\bibitem{} Hickson, P. 1990, in ``Paired and Interacting Galaxies" IAU 
Coll. 124, eds. J.W. Sulentic \& W.C. Keel, Washington, NASA,p.77
\bibitem{} Hickson, P. 1993, ApLett\&Comm 29, 1
\bibitem{} Hickson, P., Mendes de Oliveira, C., Huchra, J. P. 
\& Palumbo, G.G.C., 1992, ApJ  399, 353
\bibitem{} Hickson,  P., Kindl, E. \& Auman 1989, ApJS 70, 687
\bibitem{} Huchra, J. P., Geller, M. J., Clemens, C. M., Tokaiz, S. P. \& 
Michel, A. 1993, Harvard-Smithsonian Center for Astrophysics 
\bibitem{} Hunsberger, S. D., Charlton, J. C. \& Zaritsky, D.    
1996, ApJ 462, 50
\bibitem{} Kormendy, J.  1984, ApJ 287, 577
\bibitem{} Kormendy, J., Dressler, A., Byun, Y. Faber, S. M., Grillmair, C. 
lauer, T., Richstone, D. \& Tremaine, S. 1994, in ESO/OHP Workshop ``Dwarf 
galaxies", ed. G. Meylan \& P. Prugniel. ESO Conf. No. 49 p. 147
\bibitem{} Laurikainen, E. \&  Moles, M. 1988, AJ 96, 470
\bibitem{} Lema\^{\i}tre, G., Kohler, D., Lacroix, D., Meunier, J. P. \& 
Vin, A. 1990, A\&A 228, 546
\bibitem{} Longo, G., de Vaucouleurs, A. 1983, ``A general catalogue of
photoelectric magnitudes and colours in the UBV system of 3578 galaxies
brighter than the 16-th V-magnitude". Univ. of Texas, Austin. 
\bibitem{} Mamon, G. A. 1986, ApJ 307, 426
\bibitem{} Mamon, G. A.  1995, in ``Groups of galaxies", eds. O.G. Richter 
\& K. Borne. ASP Conf.Series Vol. 70, p 83
\bibitem{} M\'{a}rquez, I. \&  Moles, M.  1994, AJ 108, 90
\bibitem{} Mirabel, I. F. \& Wilson, A. S. 1984, ApJ 277, 92
\bibitem{} Moles, M., del Olmo A., Perea J., Masegosa J, M\'{a}rquez I. 
\& Costa V. 1994, A\&A 285, 404
\bibitem{} Molina, R., del Olmo, A., Perea, J.  \& Ripley, B.D. 1992, 
AJ 103, 666
\bibitem{} Mulchaey, J. S., Davis, D. S., Mushotzky, R. F. \& Burstein, D. 
1996, ApJ 456, 80
\bibitem{} Nieto, J., Bender, R., \& Surma, P. 1991, A\&A, 244, 37
\bibitem{} Odewah, S. C. 1996, in ``Barred Galaxies", ed. Buta, R., 
Crocker, D. A. \& Elmegreen, B. G. ASP Conference Series, Vol. 91, 30
\bibitem{} Ostriker, J.P., Lubin, L.M. \& Hernquist, L. 1995, ApJ 444, L61
\bibitem{} Pagel, B. E. J., Edmunds, M. G. \& Blackwell, D. E.  1979, MNRAS 189, 95
\bibitem{} Peletier, R. F., Valentjinm E. A., Moorwood, A. F. M. \& Freudling, 
W. 1994, A\&AS 108, 621
\bibitem{} Phookun, B., Vogel, S. N. \& Mundy, L. G. 1993, ApJ 418, 113
\bibitem{} Pickles, A. J.  1985, ApJS 59, 33
\bibitem{} Pildis, R. A. 1995, ApJ 455, 492
\bibitem{} Pildis, R. A., Bregman, J. N. \& Evrard, A. E.  1995a, ApJ 443, 514
\bibitem{} Pildis,  R. A., Bregman, J. N., \& Schombert, J. M. 1995b, 
AJ 110, 1498
\bibitem{} Pildis, R. A., Evrard, A.E.   \& Bregman, J. N.  1996, 
to appear in AJ.
\bibitem{} Ponman, T. J. \& Bertram, D. 1993, Nature 363, 51.
\bibitem{} Prandoni, I., Iovino, A. \& MacGillivray, H. T. 1994, AJ 107, 1235
\bibitem{} Ramella, M., Diaferio, A., Geller, M.J., \& Huchra, J. P. 1994, AJ 
107, 1623
\bibitem{} Rood, H.J. \& Strubble, M.F. 1994, PASP 106, 413
\bibitem{} Rood, H. J. \& Williams, B. A.  1989, ApJ 339, 772
\bibitem{} Rubin, V. C., Hunter, D. A., \& Ford, W. K. J. 1991 ApJS 76, 153
\bibitem{} Saracco, P. \& Ciliegi, P. 1995,  A\&A 301, 348
\bibitem{} Savage, B. D. \& Mathis J. S. 1979, ARA\&A 17, 73
\bibitem{} Schombert, J. M., Wallin, J. F. \& Struck-Marcell, C.  1990, AJ 99, 497
\bibitem{} Statler, T. S. 1991, AJ 102, 882
\bibitem{} Sulentic, J. 1987, ApJ 322, 605
\bibitem{} Sulentic J.W., Pietsch, W. \& Arp, H. 1995, A\&A 298, 420
\bibitem{} Sulentic J.W. \& Rabaca C.R. 1994, ApJ 429, 531.
\bibitem{} Surma, P., \&  Bender, R.  1995, A\&A 298, 405
\bibitem{} Tassi, E. \& Iovino, A.  1995, ``in Observational Cosmology: from 
galaxies to galaxy systems", in press
\bibitem{} Tikhonov, N. A. 1990,   in ``Paired and Interacting Galaxies" 
IAU Colloquium No. 124, eds. J.W. Sulentic \& W.C. Keel, NASA  p 105
\bibitem{} Tonry, J. \& Davis, M. 1979, AJ 84, 1511
\bibitem{} Toomre A. \& Toomre J., 1972 ApJ 178, 623.
\bibitem{} Unger, S. W.  et al. 1987, MNRAS 234, 745
\bibitem{} Williams, B. A. \& Rood, H. J. 1987, ApJS 63, 265
\bibitem{} Williams, B. A. \& Van Gorkom, J. H. 1995  in 
``Groups of galaxies", eds. O.G. Richter \& K. Borne. 
ASP Conf.Series Vol. 70, p 77
\bibitem{} Zepf, S. E.,  Whitmore, B. C. \& Levison, H. F. 1991, ApJ 383, 524

\end{thebibliography}
\end{document}